\documentstyle[pre,aps,graphicx,multicol,psfrag,amsmath,amssymb,citesort]{revtex}

\renewcommand{\vec}[1]{ {\bf #1} }

\newcommand{\Eq}{Eq.}
\newcommand{\Eqs}{Eqs.}

{
  {\bf ToDo: \vspace{-0.5cm}}
  \begin{list}{$\bullet$}{\setlength{\itemsep}{0pt}
  \setlength{\leftmargin}{1cm}}
}
{
  \end{list}
}

\begin{document}
\bibliographystyle{prsty}


\title{A Continuum Saltation Model for Sand Dunes}

\author{Gerd Sauermann$^{1,2}$, Klaus Kroy$^{1}$ and Hans J. Herrmann$^{1,2}$}

\address{1) PMMH, Ecole Sup\'erieure de Physique et Chimie Industrielles
  (ESPCI), 10, rue Vauquelin, 75231 Paris, Cedex 05, France}
\address{2) ICA-1, University of Stuttgart, Pfaffenwaldring 27, 70569
  Stuttgart,  Germany}



\maketitle


\vspace*{5mm}
\begin{abstract}
We derive a phenomenological continuum saltation model for aeolian sand
transport that can serve as an efficient tool for geomorphological
applications.  The coupled differential equations for the average
density and velocity of sand in the saltation layer reproduce both
known equilibrium relations for the sand flux and the time evolution
of the sand flux as predicted by microscopic saltation models. The
three phenomenological parameters of the model are a reference height
for the grain--air interaction, an effective restitution coefficient
for the grain--bed interaction, and a multiplication factor
characterizing the chain reaction caused by the impacts leading to a
typical time or length scale of the saturation transients. We determine
the values of these parameters by comparing our model with wind tunnel
measurements. Our main interest are out of equilibrium situations
where saturation transients are important, for instance at phase
boundaries (ground/sand) or under unsteady wind conditions.  We point
out that saturation transients are indispensable for a proper
description of sand flux over structured terrain, by applying the
model to the windward side of an isolated dune, thereby resolving
recently reported discrepancies between field measurements and
theoretical predictions.

\end{abstract}
\vspace*{5mm}


\begin{multicols}{2} 

\section{Introduction}

Aeolian sand transport, from the entrainment of single grains to the
formation and movement of dunes, have been studied for a long time.
One of the most important issues has been the relation $q(u_*)$
between the shear velocity $u_*$ and the saturated sand flux $q$. The
simplest flux law, which gives a cubic relation between shear velocity
and sand flux, was already introduced by Bagnold in 1936
\cite{Bagnold36}. Since that time, many new flux relations have been
proposed and used by different authors. The most important improvement
was to introduce a threshold to account for the fact that at low wind
speeds no sand transport occurs. (A overview of the historical
development can be found in Ref. \cite{Pye90}.)  One of the most
widely used flux relations with threshold was proposed by Lettau and
Lettau \cite{Lettau78}. Analytical derivations of the flux relation
starting from a microscopic picture deepened the understanding of the
aeolian transport mechanisms a lot
\cite{Owen64,Ungar87,Sorensen85,Werner90}. An application of sand flux
relations are geomorphological problems, where they are used to
calculate the erosion rate from the wind shear stress in order to
predict the evolution of a free sand surface or dune.  However, all
flux relations of the type $q(u_*)$ assume that the sand flux is
everywhere saturated.  This condition is hardly fulfilled at the
windward foot of an isolated dune, e.g. a barchan (crescent shaped
dune, discussed in Section \ref{sec:application}), where the bed
changes rapidly from bedrock to sand. Correlated measurements of the
sand flux and the wind speed performed by Wiggs \cite{Wiggs96} showed
a large discrepancy between the measured flux and theoretical
predictions of the sand flux using the relation by Lettau and Lettau
near the dune's windward foot. Numerical simulations of barchan dunes
by Wippermann and Gross \cite{Wippermann86} that employ this flux law
also revealed this problem. Apart from the conditions at the dune's
foot, it is conceivable that the sand flux may never reach saturation
on the entire windward side of a dune, where the shear velocity
increases gradually from the foot to the crest. Such effects are
obviously not captured by an equilibrium flux law. To overcome the
limitation of the equilibrium relations and to get information about
the dynamics of the aeolian sand transport, numerical simulations
based on the grain scale have been performed
\cite{Anderson88,Anderson91,McEwan91}. They showed that the typical
time to reach the equilibrium state in saltation on a flat surface is
approximately two seconds, which was later confirmed by wind tunnel
measurements \cite{Butterfield93}. The problem of simulations on the
basis of grains is that they can neither now nor in the near future be
used to calculate the evolution of macroscopic geomorphologies.

In the following we derive a dynamic continuum model that allows for
saturation transients and can thus be applied to calculate efficiently
the erosion in presence of phase boundaries and velocity gradients.
In section \ref{sec:sand_trans_salt} we introduce the phenomenology of
aeolian sand transport. In section \ref{sec:model} we develop a
continuum model for a thin fluid like sand layer on an immobile bed
including the time dependence of the sand transport and saturation
transients. The following sections discuss special cases, where
certain restrictions lead to simpler versions of the model.  In
section \ref{sec:steady_state} we discuss the saturated limit of the
model and compare it with flux relations and experimental data from
the litrature. In section \ref{sec:dyn} we disregard the spatial
dependence of the sand flux and concentrate on the time evolution of
the saltation layer. In section \ref{sec:model_geomorph} we
present a reduced ``minimal model'' that can easily be applied to
geomorphological problems. Finally, we apply this model in section
\ref{sec:application} to predict the sand flux on the central slice of
a barchan dune.

\section{Sand Transport and Saltation}
\label{sec:sand_trans_salt}

Conventionally, according to the degree of detachment of the grains
from the ground, different mechanisms of aeolian sand transport such
as suspension and bed--load are distinguished. The bed--load can be
further divided into saltation and reptation or creep. A detailed
overview of this classification can be found in Ref. \cite{Pye90}. If
we consider typical sand storms, when shear velocities are in the
range of 0.18 to 0.6$\,$m$\,$s$^{-1}$ \cite{Pye90}, particles with a
maximum diameter of 0.04--0.06$\,$mm can be transported in suspension.
The grains of typical dune sand have a diameter of the order of
0.25$\,$mm and are therefore transported via bed--load. For this reason
we neglect suspension in the following discussion. Furthermore, we do
not distinguish between saltation and reptation but consider jumping
grains with a mean trajectory length.

The saltation transport can conceptually be divided into four
sub--processes. To initiate saltation some grains have to be entrained
directly by the air. This will be called direct aerodynamic
entrainment. If there is already a sufficiently large amount of grains
in the air the direct aerodynamic entrainment is negligible and grains
are mainly ejected by impacting grains.  The entrained grains are
accelerated by the wind along their trajectory mainly by the drag
force before they impact onto the bed, again. This is called the
splash process, which comprises the complicated interaction between
bed and impacting grain and is currently the subject of theoretical
and experimental investigations \cite{Nalpanis93,Rioual2000}. Finally,
the momentum transfered from the air to the grains leads in turn to a
deceleration of the air. Through this feedback mechanism the saltation
dynamics reaches an equilibrium state, characterized by a saturated
density $\rho_s$ and an average velocity $u_s$ of the saltating
grains.

In the following we develop an effective continuum model in order to
calculate the bed--load transport. The variables will be the density
$\rho$ and mean velocity $u$ of the grains within a thin surface
layer. The closed system of equations will have three phenomenological
parameters.  The first parameter $\alpha$ models the loss of energy in
the splash process and characterizes the grain--bed interaction. It
can be thought of as an effective restitution coefficient. The second
parameter $z_1$ is a reference height between the ground and the mean
trajectory height and characterizes the air--grain interaction. These
two parameters determine the saturated sand flux $q_s$, whereas the
third parameter $\gamma$ determines the time scale $T_s$ or length
scale $l_s$ of the saturation transients.

\section{A Continuum Model for Saltation}
\label{sec:model}

We consider the bed--load as a thin fluid--like granular layer on top
of an immobile sand bed. This idea was introduced by Bouchaud et al.
\cite{Bouchaud94} and used in the following to model avalanches on
inclined surfaces near the angle of repose \cite{Mehta96,Bouchaud98}.
This general idea was later also applied to model the formation and
propagation of sand ripples \cite{Hoyle97,Hoyle99}. To avoid
cumbersome notations we restrict ourselves to a two dimensional
description of a slice of a three dimensional system that is aligned
with the wind direction. By this simplification we neglect the lateral
transport, caused for instance by gravity or diffusion, which is
typically an order of magnitude smaller than the flux in the wind
direction. A further simplification
\cite{Bouchaud94,Mehta96,Bouchaud98} is obtained by integrating over
the vertical coordinate, which leads to scalar functions and equations
for the moving layer.

We start the derivation from the mass and momentum conservation in
presence of erosion and external forces. Since the saltation layer can
exchange grains with the bed, it represents an open system with the
erosion rate $\Gamma$ as a source term,
\begin{equation}
  \frac{\partial \rho}{\partial t}
  + \frac{\partial}{\partial x} (\rho u) 
  = \Gamma.
  \label{eq:rho_s_0}
\end{equation}
Here, $\rho(x,t)$ and $u(x,t)$ denote the density and velocity of the
grains in the saltation layer, respectively. The erosion rate
$\Gamma(x,t)$ counts the number of grains per time and area that get
mobilized.

The most important forces acting on the grains are the drag force
$f_{drag}(x,t)$ when the grains are in the air, which accelerates the
grains, and the friction force $f_{bed}(x,t)$ representing the
complicated interaction with the bed, which decelerates the grains.
Thus, we can write the momentum conservation for the saltation layer
as
\begin{equation}
  \frac{\partial u}{\partial t} 
  + \left( u \frac{\partial}{\partial x} \right) u 
  = \frac{1}{\rho} \left( f_{drag} + f_{bed} \right).
  \label{eq:u_s_0}
\end{equation}
Additional forces like gravity that might be important on inclined
surfaces or diffusion caused by the splash process are neglected here
to keep the model as simple as possible, but could easily be added on
the right hand side of \Eq (\ref{eq:u_s_0}). In the following
sections we derive phenomenological expressions for the erosion rate
$\Gamma$ and the forces $f_{drag}$ and $f_{bed}$, which will finally
lead to a closed model.

\subsection{The Atmospheric Boundary Layer}

Sand transport takes place in the turbulent boundary layer of the
atmosphere, near the surface. The Navier-Stokes equation $\rho_{air}
\partial_t \vec{v} + \rho_{air} (\vec{v} \nabla)\vec{v} = -\nabla p +
\nabla \tau$, where $\vec{v}$ denotes the wind velocity, $\rho_{air}$
the density of the air, $p$ the pressure, and $\tau$ the shear stress
of the air, reduces to
\begin{equation}
  \label{eq:nav_stress}
  \frac{\partial \tau}{\partial z} = 0
  \quad \text{or} \quad
  \tau = \text{constant}
\end{equation}
by making the usual boundary layer approximation, neglecting
$\partial/\partial x, \partial/\partial y$ against $\partial/\partial
z$ and assuming steady $\partial/\partial t = 0$ and horizontally
uniform $(\vec{v} \nabla) \vec{v} = 0$ flow. At wind velocities for
which sand transport is possible, the air flow is highly turbulent.
Therefore, we can neglect the bare viscosity of the air and identify
$\tau$ with the turbulent shear stress. The standard turbulent
closure using the mixing length theory models the turbulent shear
stress $\tau$, 
\begin{equation}
  \tau = \rho_{air} \kappa^2 z^2 \left( 
    \frac{\partial v}{\partial z}
  \right)^2, 
  \label{eq:tau_dvdz}
\end{equation}
where $\kappa \approx 0.4$ denotes the von K{\'a}rm{\'a}n
constant. From equation (\ref{eq:tau_dvdz}) we obtain by introducing
the characteristic shear velocity $u_*$,
\begin{equation}
  \frac{\partial v}{\partial z} 
    = \frac{u_*}{\kappa z} 
    \quad \text{with} \quad 
    u_* = \sqrt{\frac{\tau}{\rho_{air}}}.
  \label{eq:dvdz}
\end{equation}
Integration of \Eq (\ref{eq:dvdz}) leads finally to the well
known law of the wall for turbulent flow and therefore to the
logarithmic profile of the atmospheric boundary layer,
\begin{equation}
  v(z) = \frac{u_*}{\kappa} \ln \frac{z}{z_0}
  \label{eq:v_log},
\end{equation}
where $z_0$ denotes the roughness length of the surface.

\subsection{The Grain Born Shear Stress}

In presence of saltating grains near the ground, the air can transfer
momentum to the grains, which thereby transport a part of the shear
stress down to the surface. This idea was introduced by Owen
\cite{Owen64} and has widely been used in analytical saltation models
and numerical simulations
\cite{Sorensen85,Sorensen91,Ungar87,Werner90,Anderson91,Raupach91,McEwan91}.
Accordingly, we distinguish the grain born shear stress $\tau_g$ and the
air born shear stress $\tau_a$, which together have to maintain
the overall shear stress $\tau$,
\begin{equation}
  \tau = \tau_a(z) + \tau_g(z) = \text{constant.}
  \label{eq:tau_ag}
\end{equation}
At the top $z_m$ of the saltation layer, the air born shear stress
$\tau_a$ has to be equal to the overall shear stress $\tau$,
$\tau_a(z_m) = \tau$.

The typical trajectory of a saltating grain intersects with each
height level $z$ two times, once when ascending and once when
descending. Between these two intersections the wind accelerates the
grain along its trajectory, thereby increasing its horizontal velocity
between the ascending and descending intersection. From this velocity
difference we can calculate the shear stress transported by the grains
at each level $z$,
\begin{equation}
  \tau_g(z) = \Phi [u_{down}(z) - u_{up}(z)]
            = \Phi \Delta u_g(z),
  \label{eq:tau_updown}
\end{equation}
where $\Phi$ denotes the flux of grains impacting onto the surface,
and $u_{up}$ and $u_{down}$ the horizontal velocity of the grains at
the ascending and descending intersection of the trajectory with the
height level $z$. 

\begin{figure}[tb]
  \begin{center}
    \psfrag{length}[c][c]{$l$}
    \psfrag{width}[c][c]{$a$}
    \includegraphics[width=0.9\columnwidth]{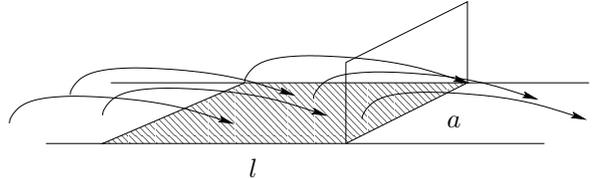}
  \end{center}
  \caption{Sketch of the horizontal sand flux $q$ caused by
    saltating grains passing the vertical rectangle. The dashed
    rectangle shows the surface area $l$ times $a$ that is used to
    calculate the flux $\Phi$ of grains impacting onto the surface,
    where $l$ is the length of the saltation trajectory and $T$ the
    flight time of the grain.}
  \label{fig:sketch_saltation}
\end{figure}

As sketched in Fig. \ref{fig:sketch_saltation} we can relate both the
horizontal sand flux $q$ and the flux of grains impacting onto the surface
$\Phi$ to the number of grains $N$, their mass $m$, the saltation
length $l$, and the flight $T$ time for the trajectory,
\begin{equation}
  \label{eq:q_phi_N}
  \Phi = \frac{N \, m}{l \, a \, T} = \frac{q}{l},
\end{equation}
where $a$ is the arbitrary width of the slice, considered here.
Using \Eq (\ref{eq:tau_updown}) and (\ref{eq:q_phi_N}) the grain
born shear stress can be expressed by the horizontal flux of grains,
\begin{equation}
  \label{eq:tau_g_of_q}
  \tau_g(z) = \frac{q}{l} \Delta u_g(z).
\end{equation}

In the following we restrict the discussion to the grain born shear
stress $\tau_{g0}$ at the ground that is given by the momentum
transfer from the grains to the bed during their impacts as sketched
in Figure~\ref{fig:sketch_saltation}. We denote quantities that are
taken at the ground with an index 0, e.g. $\tau_{g0} = \tau_g(z_0)$,
where $z_0$ is the height at the ground. We need the impact and
ejection velocities of the grains or moreover the change in horizontal
velocity $\Delta u_{g0}$ at the ground to calculate the momentum
transfer. In order to keep the discussion simple we will directly
formulate the model in terms of mean values for the trajectory length
$l$ and its time $T$ instead of writing everything in terms of not
well known distribution functions.

The average saltation length $l$ and the average saltation time $T$
are related by $l=u \, T$. We estimate both as the flight time and
length of a simple ballistic trajectory,
\begin{equation}
  T = \frac{2 u_{z0}}{g},
  \quad \quad
  l = u \frac{2 u_{z0}}{g},
  \label{eq:traj_DT_l}
\end{equation}
where $u_{z0}$ is the vertical component of the initial velocity of
the grain and $g$ the acceleration by gravity. The typical values for
the time and length of a mean saltation trajectory depend on the shear
velocity $u_*$, $T(u_*) \approx 1.7 \, u_*/g$ and $l(u_*) \approx 18
\, u_*^2/g$ \cite{Nalpanis93}. For a shear velocity
$u_*=0.5\,$m$\,$s$^{-1}$ we obtain $T \approx 0.08\,$s and $l \approx
0.45\,$m. Calculations of grain trajectories confirm that this
approximation gives the correct order of magnitude for the flight time
$T$, e.g. S{\o}rensen obtained $T=1.75 u_*/g$ \cite{Sorensen91}.
Inserting the saltation length of \Eq (\ref{eq:traj_DT_l}) in
\Eq (\ref{eq:tau_g_of_q}) and using $q=\rho u$ we obtain,
\begin{equation}
  \tau_{g0} = \rho \, \frac{g}{2} \, \frac{\Delta u_{g0}}{u_{z0}}.
  \label{eq:tau_s_micro}
\end{equation}
The relation between the impact and ejection velocity is in general
defined by the splash function
$S(u_{im},\alpha_{im},...;u_{ej},\alpha_{ej},...)$ giving the
probability that a grain is ejected at a certain angle and velocity
due to an impacting grain with a certain angle and velocity
\cite{Anderson91,Nalpanis93,Rioual2000}. The simplest possibility is
to take the vertical component of the ejection velocity $u_{z0}$
proportional to the horizontal velocity difference $\Delta u_{g0}$ and
to neglect the angle dependence,
\begin{equation}
  u_{z0} = \alpha \, \Delta u_{g0}.
  \label{eq:splash_delta}
\end{equation}
Here, we have introduced the first model parameter $\alpha$ that
represents an effective restitution coefficient for the grain--bed
interaction. In principle, it can be calculated from the splash
function $S$, but here we regard it as a phenomenological parameter to
be determined later by comparing the model with experimental results.
With \Eq (\ref{eq:splash_delta}), \Eq (\ref{eq:tau_s_micro})
reduces to the simple result
\begin{equation}
  \tau_{g0} = \rho \frac{g}{2 \alpha}
  \label{eq:tau_s}
\end{equation}
for the grain born shear stress. The fact that it is linear in the
density $\rho$ and independent of the velocity $u$ is an interesting
result that is not a priori obvious. Indeed, if we had chosen the
effective restitution coefficient $\alpha(u)$ as a function of the
velocity we would obtain a velocity dependence. Why is the velocity
not or only of little importance for $\tau_{g0}$? This can be
explained by looking at the trajectory length of the saltating grains,
which scales with the square of the horizontal velocity ($l \propto
u^2$) leading to a decrease of the density of impacts with increasing
$u$. This compensates the higher momentum transfer of a single grain
due to its higher velocity.

\subsection{Erosion and Deposition Rates}

Grains impacting onto a bed of grains either rebounce directly or
remain on the bed. In the latter case the energy of the impact can
dislodge several new grains leading to a strong amplification of the
grain density in the saltation layer and in turn to erosion.  The
probability that a certain number of grains leaves the surface with a
certain velocity is in general given by the splash
function $S$ \cite{Anderson91,Nalpanis93,Rioual2000}, mentioned above.
A part of the information contained in the splash function has already
been comprised into the model parameter $\alpha$ that relates the
average horizontal velocity difference to the vertical component of
the average ejection velocity. Another part can be represented by the
average number $n$ of grains disloged by an impacting grain. Together,
$\alpha$ and $n$ determine the simple effective splash function used
in this model,
\begin{equation}
  \label{eq:splash_simple}
  S(u_{z0},\Delta u_{g0}) = n \, \delta(\alpha \, \Delta u_{g0} - u_{z0}).
\end{equation}
Using the average number $n$ of grains dislodged by an impacting grain
we can define the erosion rate as the net average rate of grains
leaving the surface, which is the difference between the flux of
grains leaving the surface and the flux of grains impacting onto it,
\begin{equation}
  \Gamma = \Phi ( n - 1 )
         = \frac{\tau_{g0}}{\Delta u_{g0}} \left( n-1 \right).
  \label{eq:gamma_s_dyn}
\end{equation}
It is important to realize that the air born shear stress within the
saltation layer, and therewith $n$, is lowered if the number of grains
in the saltation layer increases and vice versa. This is the feedback
effect discussed in section \ref{sec:sand_trans_salt}. According to
Owen \cite{Owen64}, the air born shear stress at the bed $\tau_{a0}$
in the equilibrium is just large enough to keep saltation alive and
therefore close to the threshold $\tau_t=\rho_{air} u_{*t}^2$. For
$\tau_{a0} > \tau_t$, the number of grains in the saltation layer
increases in a chain reaction ($n>1$) whereas for $\tau_{a0} < \tau_t$
($n<1$) saltation cannot be maintained. To model the average number
$n$ of dislodged grains out of equilibrium we write $n$ as a function
$n(\tau_a/\tau_t)$ with $n(1)=1$. We furthermore assume that $n$ can
be expanded into a Taylor series at the threshold and neglect all
terms after the linear order,
\begin{equation}
  n \left( \frac{\tau_{a0}}{\tau_t} \right)
    = 1 + \tilde{\gamma} \left( \frac{\tau_{a0}}{\tau_t} - 1 \right) ...
  \label{eq:ampl_n}
\end{equation}
The model parameter $\tilde{\gamma}$ characterizes the strength of the
erosion and determines how fast the system reaches the equilibrium or
reacts to perturbations. It depends on microscopic quantities such as
the time of a saltation trajectory or the grain--bed interaction,
which are not available in the scope of this model. Therefore, we
have to determine $\tilde{\gamma}$ later by comparison with
measurements or microscopic computer simulations.

If we insert \Eq (\ref{eq:tau_ag}) and (\ref{eq:ampl_n}) in
\Eq (\ref{eq:gamma_s_dyn}), we obtain for the erosion rate
\begin{equation}
  \Gamma = \tilde{\gamma} \frac{\tau_{g0}}{\Delta u_{g0}}
        \left( \frac{\tau - \tau_{g0}}{\tau_t} - 1 \right).
  \label{eq:Gamma_s_delta_u}
\end{equation}
Assuming that the difference between impact and eject velocity of the
grains is proportional to the mean grain velocity ($\Delta
u_{g0} \propto u$) finally leads to 
\begin{equation}
  \Gamma = \gamma \frac{\tau_{g0}}{u}
        \left( \frac{\tau - \tau_{g0}}{\tau_t} - 1 \right),
  \label{eq:Gamma_s}
\end{equation}
where the proportionality constant is incorporated in $\gamma$. In
order to close the system of equations we have to insert the 
expression for the grain born shear stress $\tau_{g0}$ from \Eq
(\ref{eq:tau_s}).

Up to now, we discussed the entrainment of grains due to impacts of
other grains, but if the air shear stress exceeds a certain value
$\tau_{ta} > \tau_t$, called aerodynamic entrainment threshold, grains
can directly be lifted from the bed. These directly entrained grains
have been neglected up to now, because they are only important to
initiate saltation \cite{Anderson88,Anderson91}. Anderson
\cite{Anderson91} proposed an aerodynamic entrainment rate
proportional to the difference between the air born shear stress
$\tau_a$ and the threshold $\tau_{ta}$,
\begin{equation}
  \label{eq:gamma_a}
  \Gamma_a = \Phi_a \left( \frac{\tau_{a0}}{\tau_{ta}} - 1 \right) 
           = \Phi_a \left( \frac{\tau - \tau_{g0}}{\tau_{ta}} - 1 \right),
\end{equation}
where $\Phi_a \approx$~5.7~10$^{-4}\,$kg$\,$m$^{-2}\,$s$^{-1}$
\cite{Anderson91} is a model parameter defining the strength of the
erosion rate for aerodynamic entrainment.  This formula for the direct
aerodynamic entrainment rate $\Gamma_a$ has a similar structer as
\Eq (\ref{eq:Gamma_s}) for saltation induced entrainment, but the
prefactors are different.

\subsection{Forces}

In section \ref{sec:model} we introduced the drag and friction forces,
$f_{drag}$ and $f_{bed}$, acting on the saltation layer. In the
following we have to specify these forces. Modeling the friction force
is simple, because we already derived an expression for the grain born
shear stress at the ground $\tau_{g0}$, \Eq (\ref{eq:tau_s}). The
bed friction $f_{bed}$ has exactly to compensate this grain born shear
stress,
\begin{equation}
  \label{eq:f_bed}
  f_{bed} = - \tau_{g0}.
\end{equation}
We represent the wind force acting on the grains inside the saltation
layer by the Newton drag force,
\begin{equation}
  F_{drag} = \frac{1}{2} \rho_{air} C_d \frac{\pi d^2}{4} 
        (v_{air} - v_g)|v_{air} - v_g|,
  \label{eq:drag_part}
\end{equation}
of a spherical particle, where $d$ denotes the grain diameter, $C_d$ the
drag coefficient, $v_g$ the velocity of a grain, and $v_{air}$ the
velocity of the air.  Multiplying $F_{drag}$ with the density $\rho$ of
the saltation layer and dividing it by the mass $m$ of a grain with
diameter $d$ and density $\rho_{quartz}$, we obtain the drag force
acting on a volume element of the saltation layer,
\begin{equation}
  f_{drag} = \rho \frac{3}{4} C_d \frac{\rho_{air}}{\rho_{quartz}} 
             \frac{1}{d} (v_{eff} - u) |v_{eff} - u|.
  \label{eq:f_drag}
\end{equation}
Here, $v_{eff}$ is an effective wind velocity, which is the wind speed
taken at a reference height $z_1$ within the saltation layer. This
reference height is another model parameter and has to be determined
by comparing the sand flux with measured data as we will do it in
section \ref{sec:steady_state}. Neglecting the effect of the saltating
grains on the wind field, we could use the logarithmic profile at
$z_1$ to calculate the effective wind speed. But saltating grains in
the air change the air shear stress and the wind speed. This is the
feed back effect mentioned above and the mechanism how an equilibrium
sand flux is reached. In order to calculate the perturbed wind speed
we use again the standard turbulent closure (\ref{eq:dvdz}), which
relates the strain rate to the turbulent shear stress. In contrast to
the case without grains, where the shear stress $\tau$ of the air is
constant in $z$, the air born shear stress $\tau_a$ as given by
\Eq (\ref{eq:tau_ag}) is now varying in $z$,
\begin{equation}
  \frac{\partial v}{\partial z} 
    = \frac{u_*}{\kappa z} \sqrt{1 - \frac{\tau_g(z)}{\tau}}.
  \label{eq:dvdz_veff1}
\end{equation}
The profile of $\tau_g(z)$ was found to be nearly exponential in
simulations \cite{Anderson91}.  Furthermore, we already know the grain
born shear stress at the ground $\tau_{g0}$. Hence, we can write
\begin{equation}
  \frac{\partial v}{\partial z} 
    = \frac{u_*}{\kappa z} \sqrt{1 - \frac{\tau_{g0}}{\tau} e^{-z/z_m}},
  \label{eq:dvdz_veff2}
\end{equation}
where $z_m$ denotes the mean saltation height. The integration of
\Eq (\ref{eq:dvdz_veff2}) has to be performed from the roughness
height $z_0$ to a reference height $z_1 < z_m$. Therefore we
linearize the exponential function to integrate \Eq
(\ref{eq:dvdz_veff2}) and obtain the effective wind velocity
%
\begin{multline}
  \label{eq:veff1}
  v_{eff} = 
  \frac{u_*}{\kappa} \sqrt{1-\frac{\tau_{g0}}{\tau}} \\
    \left( 
      2 A_1 - 2 A_0 + \ln \frac{(A_1-1)(A_0+1)}{(A_1+1)(A_0-1)}
    \right)
\end{multline}
%
with
\begin{equation}
  \label{eq:veff_terms}
  A_i = \sqrt{1+\frac{z_i}{z_m} \frac{\tau_{g0}}{\tau-\tau_{g0}}} \; .
\end{equation}
For a reference height $z_1$ much smaller than the mean saltation
height $z_m$ ($z_0 < z_1 \ll z_m$) we can simplify \Eq
(\ref{eq:veff1}) to
\begin{equation}
  \label{eq:v_eff_of_tau_g0}
  v_{eff} = 
  \frac{u_*}{\kappa} \sqrt{1-\frac{\tau_{g0}}{\tau}} 
    \left(  
      2 A_1
      - 2 + \ln \frac{z_1}{z_0} 
    \right) .
\end{equation}
For vanishing grain born shear stress $\tau_{g0} \rightarrow 0$, the
effective wind velocity $v_{eff}$ reduces to the velocity of the
undisturbed logarithmic profile at height $z_1$. The values of the
parameters $z_0$ and $z_m$ can be obtained from measurements, whereas
the value of the reference height $z_1$ is a free phenomenological
parameter of the model that we have to estimate by comparing the
saturated flux predicted by our model with measurements.

\subsection{Closed Model}
\label{sec:closed_model}

So far, we have introduced the erosion rate $\Gamma$, the drag force
$f_{drag}$, and the bed interaction $f_{bed}$ and expressed them in
the preceding sections in terms of the mean density $\rho$ and the
mean velocity $u$ of the saltating grains. Furthermore, we have
introduced two model parameters $\alpha$ and $z_1$ determining the
equilibrium state of the saltation layer, and the parameter $\gamma$
that controls the relaxation to equilibrium. Inserting \Eqs
(\ref{eq:Gamma_s}) and (\ref{eq:tau_s}) in \Eq (\ref{eq:rho_s_0})
leads to an equation for the sand density $\rho$ in the
saltation layer,
\begin{equation}
  \label{eq:dt_rho_1}
  \frac{\partial \rho}{\partial t}
  + \frac{\partial}{\partial x} (\rho u) 
  = \gamma \frac{g}{2 \alpha} \frac{\tau-\tau_t}{\tau_t}
  \frac{\rho}{u} \left(
    1-\frac{g}{2 \alpha} \frac{1}{\tau-\tau_t} \rho
  \right).
\end{equation}
Here, we can identify two important physical quantities, the saturated
density $\rho_s$ and the characteristic time $T_s$ that define the
steady state and the transients of the sand density $\rho$,
respectively,
\begin{equation}
  \label{eq:rho_s_tau}
  \rho_s = \frac{2 \alpha}{g} \left( \tau - \tau_t \right),
\end{equation}
\begin{equation}
  \label{eq:T_s}
  T_s = \frac{2 \alpha u}{g} \, \frac{\tau_t}{\gamma (\tau - \tau_t)}.
\end{equation}
With these expressions we can rewrite \Eq (\ref{eq:dt_rho_1}) in
a more compact form,
\begin{equation}
  \label{eq:dt_rho}
  \frac{\partial \rho}{\partial t}
  + \frac{\partial}{\partial x} (\rho u) 
  = \frac{1}{T_s} \rho \left( 1 - \frac{\rho}{\rho_s} \right).
\end{equation}
Direct aerodynamic entrainment, i.e. the initiation of saltation, has
been neglected in \Eq (\ref{eq:dt_rho}), but can easily be
included by adding the erosion rate $\Gamma_a$ of \Eq
(\ref{eq:gamma_a}) to the right hand side. Furthermore, inserting
\Eqs (\ref{eq:f_bed}) and (\ref{eq:f_drag}) in \Eq
(\ref{eq:u_s_0}) leads to a model for the sand velocity $u$ in
the saltation layer,
%
\begin{multline}
  \label{eq:dt_u}
  \frac{\partial u}{\partial t} 
  + \left( u \frac{\partial}{\partial x} \right) u 
  = \\
  \frac{3}{4} C_d \frac{\rho_{air}}{\rho_{quartz}} \frac{1}{d} 
    (v_{eff} - u)|v_{eff} - u|
    - \frac{g}{2 \alpha},
\end{multline}
%
with $v_{eff}$ defined in (\ref{eq:v_eff_of_tau_g0}). Finally, the
\Eqs (\ref{eq:dt_rho}), (\ref{eq:rho_s_tau}), (\ref{eq:T_s}), 
(\ref{eq:dt_u}), and (\ref{eq:v_eff_of_tau_g0}) define the closed
model for the sand flux in the saltation layer.

We want to emphasize that $T_s(\tau,u)$ and $l_s(\tau,u) = T_s \, u$ are not
constant, but depend on the external shear stress $\tau$ of the wind
and the mean grain velocity $u$. Using \Eq (\ref{eq:traj_DT_l})
we can relate the characteristic time $T_s$ and length $l_s$ of the
saturation transients to the saltation time $T$ and the saltation
length $l$ of the average grain trajectory,
\begin{equation}
  \label{eq:T_s_of_T}
  T_s = T \, \frac{\tau_t}{\tilde{\gamma} (\tau - \tau_t)},
  \quad \quad
  l_s = l \, \frac{\tau_t}{\tilde{\gamma} (\tau - \tau_t)}.
\end{equation}
For typical wind speeds, the time to reach saturation is in the order
of 2$\,$s \cite{Anderson88,Anderson91,McEwan91}. Assuming a grain
velocity of 3--5$\,$m$\,$s$^{-1}$ \cite{Willetts85} we obtain a length
scale of the order of 10$\,$m for saturation. This length scale is
large enough to play an important role in dune formation. 
We want to emphasize that the characteristic length $l_s$ of the
saturation transients, \Eq (\ref{eq:T_s_of_T}), naturally result
from the saltation kinetics (in contrast to the heuristic ``adaptation
length'' of Ref. \cite{Dijk99}). Their functional dependence can be
interpreted in the following way. The dominant mechanism to adapt the
grain density in the saltation layer due to a change in external
conditions is by the chain reaction process modeled in \Eq
(\ref{eq:ampl_n}). Hence, $l_s$ and $T_s$ depend inversely onto the
``stiffness'' $\tilde \gamma (\tau / \tau_t -1)$ of this multiplication
process. Since the average grain can only influence the number of
grains in the saltation layer at the discrete times and positions
where it interacts with the bed, the saturation time/length are
proportional to the time/length of the average trajectory. The
resulting non-trivial dependence of the saturation kinetics on the
external conditions, which may be appreciated from Figure \ref{fig:ls}
in Section \ref{sec:model_geomorph}, is essential for a proper
description of aeolian sand transport in structured terrain.

The mass and momentum \Eqs (\ref{eq:dt_rho}) and (\ref{eq:dt_u})
are coupled partial differential equations and difficult to solve. In
the following sections we will first discuss the fully saturated
situation and later some dynamical properties of the saltation layer.
Finally, we simplify the model in order to arrive at a minimal
model that can easily be applied to geomorphological problems.

\section{Saturated Flux}
\label{sec:steady_state}

The full dynamics of the saltation layer must be evaluated
numerically, whereas the saturated flux --- the stationary solution
($\partial/\partial t = 0$) for a constant external shear stress
($\tau(x,t) = \tau$) and a homogeneous bed ($\partial/\partial x = 0$) ---
can be calculated analytically from \Eq (\ref{eq:dt_rho_1}) and
(\ref{eq:dt_u}). For shear velocities below the threshold ($u_* <
u_{*t}$) the solution is trivial,
\begin{equation}
  \label{eq:below_threshold}
  \rho_s(u_*) = 0, \quad u_s(u_*) = 0, \quad \text{and} 
  \quad q_s(u_*) = 0.
\end{equation}
Above the threshold ($u_* >u_{*t}$) we obtain from \Eq
(\ref{eq:rho_s_tau}) for the steady state density $\rho_s$,
\begin{equation}
  \rho_s(u_*) = 
  \frac{2 \alpha \rho_{air}}{g} (u_*^2 - u_{*t}^2).
  \label{eq:rho_s_steady}
\end{equation}
Likewise we obtain from \Eqs (\ref{eq:dt_u}) the steady state
velocity $u_s$,
%
\begin{multline}
  u_s(u_*) = \\
  \frac{2 u_*}{\kappa} 
    \sqrt{\frac{z_1}{z_m} + \left(1 - \frac{z_1}{z_m}\right) \frac{u_{*t}^2}{u_*^2}} 
  - \frac{2 u_{*t}}{\kappa}
  + u_{st} ,
  \label{eq:u_s_steady}
\end{multline}
%
where
\begin{equation}
  \label{eq:u_min}
  u_{st} \equiv u_s(u_{*t}) = 
           \frac{u_{*t}}{\kappa} \ln \frac{z_1}{z_0} -
            \sqrt{\frac{2 \, g \, d \, \rho_{quartz}}
              {3 \, \alpha \, C_d \, \rho_{air}}
            }.
\end{equation}
is the minimum velocity of the grains in the saltation layer occurring
at the threshold. In contrast to the density $\rho_s$ the velocity
$u_s$ does not go continuously to zero near the threshold $u_{*t}$.
This is intuitively obvious, because grains at the threshold have
already a finite velocity $u_{st}$. Finally, we can write for the
steady state flux $q_s = u_s \rho_s$,
\begin{multline}
  q_s =
    2 \alpha \frac{\rho_{air}}{g} \left(u_*^2 - u_{*t}^2 \right)\\
       \left( u_* \frac{2}{\kappa} 
         \sqrt{\frac{z_1}{z_m} + \left(1 - \frac{z_1}{z_m}\right) \frac{u_{*t}^2}{u_*^2}} 
         - \frac{2 u_{*t}}{\kappa}
         + u_{st}
       \right).
  \label{eq:q_s_steady}
\end{multline}
For large wind speeds ($u_* \gg u_{*t}$) the flux is asymptotically
proportional to $u_*^3$, which is in accord with the predictions of
other saltation models \cite{Bagnold41,Lettau78,Sorensen91}.

The saturated flux $q_s$, given by \Eq (\ref{eq:q_s_steady}), is
now used to determine the model parameters $\alpha$ and $z_1$ by
fitting it to flux data measured in a wind tunnel by White
\cite{White91}.  Using the literature values: $g=9.81\,$m$\,$s$^{-2}$,
$\rho_{air}=1.225\,$kg$\,$m$^{-3}$,
$\rho_{quartz}=2650\,$kg$\,$m$^{-3}$, $z_m=0.04\,$m,
$z_0=2.5~10^{-5}\,$m, $D=d=250\,\mu$m, $C_d=3$ and
$u_{*t}=0.28\,$m$\,$s$^{-1}$ \cite{Owen64,Pye90,Anderson91} we obtain
for the two model parameters $\alpha=0.35$ and $z_1=0.005\,$m. 
For comparison we fitted to the same set of data the sand transport
laws given by Bagnold \cite{Bagnold41},
\begin{equation}
  \label{eq:q_bagnold}
  q_B = C_B \frac{\rho_{air}}{g} \sqrt{\frac{d}{D}} u_*^3,
\end{equation}
Lettau and Lettau \cite{Lettau78},
\begin{equation}
  \label{eq:q_lettau}
  q_L = C_L \frac{\rho_{air}}{g} u_*^2 (u_*-u_{*t}),
\end{equation}
and S{\o}rensen \cite{Sorensen91},
\begin{equation}
  \label{eq:q_sorensen}
  q_S = C_S \frac{\rho_{air}}{g} u_* (u_* - u_{*t}) (u_* + 7.6*u_{*t} + 205)
\end{equation}
and obtained $C_B=1.98$, $C_L=4.10$, and $C_S=0.011$.

\begin{figure}[tb]
  \begin{center}
    \psfrag{norm_flux}[c][c]{flux $q$/$q_0$}
    \psfrag{u_star}[t][c]{shear velocity $u_*$/$u_{*t}$}
    \psfrag{Sauermann}[r][r]{\tiny This model}
    \psfrag{Bagnold}[r][r]{\tiny Bagnold}
    \psfrag{Lettau}[r][r]{\tiny Lettau}
    \psfrag{Sorensen}[r][r]{\tiny S{\o}rensen}
    \psfrag{White}[r][r]{\tiny White}
    \includegraphics[width=0.95\columnwidth]{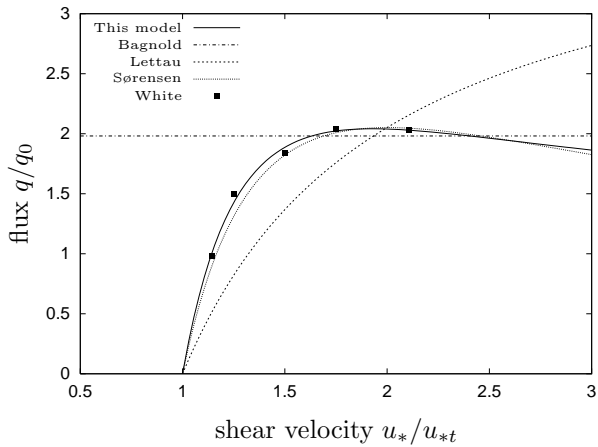}
    \vspace*{4mm}
    \caption{Comparison of the different theoretical flux relations
      (\ref{eq:q_s_steady})-(\ref{eq:q_sorensen}) fitted to wind
      tunnel data of White. The fluxes are normalized by
      $q_0=\rho_{air} / (g u_*^3)$. The saturated flux of our model
      and the relation of S{\o}rensen reproduce quite well the data,
      whereas the others do not show the same structure.}
    \label{fig:q_steady_state}
  \end{center}
\end{figure}

The results are shown in Figure~\ref{fig:q_steady_state}, where the
flux normalized by $q_0 = \rho_{air} / (g u_*^3)$ is plotted versus
the normalized shear velocity $u_*/u_{*t}$. Our result resembles
S{\o}rensen's equation but differs from the flux relation given by Lettau and
Lettau.  For high shear velocities all transport laws show a cubic
dependence on the shear velocity.

The comparison of the saturated sand flux with experimental data
determined the two phenomenological parameters $\alpha$ and $z_1$. As
anticipated above, the value obtained for $z_1$, the reference
height for momentum transfer, is well below $z_m$, the mean
height of the saltation trajectories. This justifies a posteriori the
linearization in \Eq (\ref{eq:dvdz_veff2}).

\section{Dynamics}
\label{sec:dyn}

\begin{figure}[tb]
  \begin{center}
    \psfrag{flux}[b][c]{$q$ in kg~m$^{-1}$~s$^{-1}$}
    \psfrag{time}[t][c]{$t$ in s}
    \psfrag{u03}[r][r]{\tiny $u_*=0.3$}
    \psfrag{u04}[r][r]{\tiny $u_*=0.4$}
    \psfrag{u05}[r][r]{\tiny $u_*=0.5$}
    \psfrag{u06}[r][r]{\tiny $u_*=0.6$}
    \psfrag{u07}[r][r]{\tiny $u_*=0.7$}
    \includegraphics[width=0.95\columnwidth]{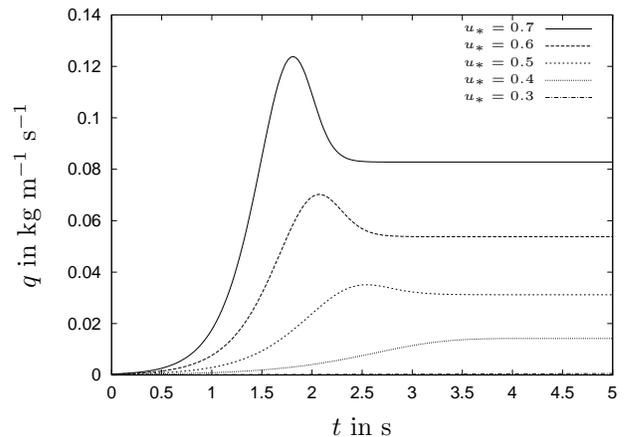}
    \vspace*{4mm}
    \caption{Numerical simulations of the time evolution of the full
      model given by \Eq (\ref{eq:dt_rho}) and (\ref{eq:dt_u})
      with a constant shear velocity $u_*$.}
    \label{fig:q_trans}
  \end{center}
\end{figure}

After the saturated case has been studied in the preceding section, we
now investigate the dynamics of the saltation layer in order to get an
estimate for the saturation time $T_s$ and thus for the model
parameter $\gamma$.  Figure~\ref{fig:q_trans} shows numerical
solutions of \Eq (\ref{eq:dt_rho}) and (\ref{eq:dt_u}) for the
time evolution of the sand flux $q=u \, \rho$ using spatially
homogeneous conditions ($\partial / \partial x = 0$).  To get rid of
the free parameters in \Eq (\ref{eq:gamma_a}) we neglected
$\Gamma_a$, thus disregarding direct aerodynamic entrainment, and
assumed instead a small initial density. Due to the multiplication
effect of the saltation process, the flux increases first
exponentially and reaches the equilibrium state after passing through
a maximum at $t \approx 2\,$s. The time transients are controlled by
the parameter $\gamma$ and compare well with measurements by
Butterfield \cite{Butterfield93} and microscopic simulations by
Anderson \cite{Anderson88,Anderson91} and McEwan \cite{McEwan91} for
$\gamma \approx 0.4$. An important result of the simulations of
Anderson was the dependence of the saturation time on the shear
velocity and the overshoot near $t \approx 2\,$s. Both features are
well reproduced by our model.

\section{A Minimal Model for Geomorphological Applications}
\label{sec:model_geomorph}

The change of desert topographies, e.g. the movement, growth, and
shrinkage of dunes, depends mainly on the sand transport or erosion
and the perturbations of the wind field caused by the topographies
themselves. Here, we restrict ourselves to the problem of the sand
transport and assume that the shear velocity above a certain
topography is known. The time evolution of the topography $h(x,t)$
is then given by the mass conservation,
\begin{equation}
  \label{eq:dhdt}
  \frac{\partial h}{\partial t} = - \frac{1}{\rho_{sand}}
      \frac{\partial q}{\partial x},
\end{equation}
where $\rho_{sand}$ is the mean density of the immobile dune sand. To
obtain the sand flux $q(x,t)$ one can in principle solve the coupled
differential equations for the density $\rho$, \Eq
(\ref{eq:dt_rho}), and velocity $u$, \Eq (\ref{eq:dt_u}), of the
sand in the saltation layer. However, for most geomorphological
applications a simplified version of our model will be sufficient. In
the following, we first derive this ``minimal model'' from the
\Eqs given in Section~\ref{sec:closed_model} and then show its
usefulness by discussing a particular practical application in
Section~\ref{sec:application}.

The first simplification is to use the stationary solution
($\partial/\partial t=0$) of \Eq (\ref{eq:dt_rho}) and
(\ref{eq:dt_u}). This can be justified by the fact that there are
several orders of magnitude between the time scale of saltation
(approximately 2s) and the time scale of the surface evolution of
a dune (several days or weeks). 

\begin{figure}[tb]
  \begin{center}
    \psfrag{ls}[b][c]{$l_s$ in m}
    \psfrag{u_star}[t][c]{$u_*/u_{*t}$}
    \includegraphics[width=0.95\columnwidth]{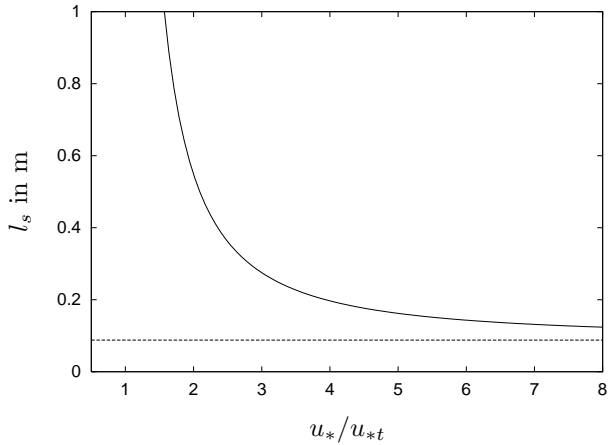}
    \vspace*{4mm}
    \caption{The saturation length $l_s$ from \Eq
      (\ref{eq:l_s_of_u_s}) is asymptotically constant for high shear
      velocities (horizontal line), but diverges for shear velocities
      near the threshold.}
    \label{fig:ls}
  \end{center}
\end{figure}

Next, we consider the convective term ($u \partial_x u$) that is only
important at places where large velocity gradients occur. This is for
instance the case in the wake region behind the brink of a dune, where
the wind speed at the ground decreases drastically due to the
flow--separation of the air. Here, the inertia of the grains becomes
important. To solve the model for the deposition in a wake region we
want to consider an idealized brink situation, where both the wind speed
and the friction with the bed drop discontinuously from a finite
value to zero. In this case, \Eq (\ref{eq:dt_u}) reduces to
\begin{equation}
  \label{eq:u_stat_lee}
  \frac{1}{2} \frac{\partial u^2}{\partial x} 
  = \frac{3}{4} C_d \frac{\rho_{air}}{\rho_{quartz}} \frac{1}{d} u^2 .
\end{equation}
This predicts an exponential decrease of the grain velocity over a
characteristic length scale $l_{dep} = 4 \, d \, \rho_{quartz} / (3 \,
C_d \, \rho_{air}) \approx 0.25\,$m.  Hence, the deposition takes
place within a length $l_{dep} \ll l_s$ much shorter than the
saturation length $l_s$ on the windward side. Field measurements of
lee side deposition agree with this conclusion \cite{Richard95}.
Outside the wake regions, on the other hand, we can neglect the
convective term ($u \partial_x u$) and we obtain the mean stationary
grain velocity from \Eq (\ref{eq:dt_u})
\begin{equation}
  \label{eq:u_of_rho_stationary}
  u(\rho) = v_{eff}(\rho) - \sqrt{\frac{2 \, g \, d \, \rho_{quartz}}
                                       {3 \, \alpha \, C_d \,
                                         \rho_{air}}
                            },
\end{equation}
where $v_{eff}(\rho)$ is defined by \Eq
(\ref{eq:v_eff_of_tau_g0}) and (\ref{eq:tau_s}). Furthermore, near the
threshold $\tau_t$ we can approximate $v_{eff}(\rho)$ by
$v_{eff}(\rho_s)$ making only a negligible error. For high shear
stresses this is not in general possible. But, the sand flux in
macroscopic geomorphological applications is nearly everywhere
saturated, apart from places where external variables change
discontinously, e.g.  at a phase boundary bedrock/sand or at a
flow--separation (see Fig.  \ref{fig:flux_shear}). Hence, we can
replace the density $\rho$ by the saturated density $\rho_s$ for most
applications, where the exact value of the flux at these places is not
of importance. The advantage of this approximation is that the
velocity $u$ is decoupled from the density $\rho$ and we can insert
the saturated velocity $u_s$ of \Eq (\ref{eq:u_s_steady}) in
\Eq (\ref{eq:dt_rho}). Rewriting \Eq (\ref{eq:dt_rho}) in
in terms of the sand flux $q=\rho \, u_s$ and the saturated sand flux
$q_s = \rho_s \, u_s$ of \Eq (\ref{eq:q_s_steady}) leads to an
equaation for the sand flux $q$,
\begin{equation}
  \label{eq:q_stationary}
  \frac{\partial}{\partial x} q = 
    \frac{1}{l_s} q \left( 1 - \frac{q}{q_s} \right),
\end{equation}
where 
\begin{equation}
  \label{eq:l_s_of_u_s}
    l_s = \frac{2\alpha}{\gamma} 
          \frac{u_s^2}{g} \, 
          \frac{\tau_t}{\tau - \tau_t}
\end{equation}
is the saturation length depicted in Figure \ref{fig:ls}.

We want to emphasize that our most important result for
geomorphological applications is \Eq (\ref{eq:q_stationary}),
which extends a common saturated flux law by incorporating saturation
transients. It is to some extent independent of the particular form of
the functions $l_s$ and $q_s$ given in \Eq (\ref{eq:l_s_of_u_s})
and (\ref{eq:q_s_steady}). The latter can be regarded as additional
predictions of our model which one could also replace by other
phenomenological relations or data tables obtained from wind tunnel
measurements if this turns out to be more suitable.

\begin{figure}[tb]
  \begin{center}
    \psfrag{flux}[b][c]{$q$ in kg~m$^{-1}$~s$^{-1}$}
    \psfrag{distance}[t][c]{$x$ in m}
    \psfrag{ustar0.3}[r][r]{\tiny $u_*=0.3$}
    \psfrag{ustar0.4}[r][r]{\tiny $u_*=0.4$}
    \psfrag{ustar0.5}[r][r]{\tiny $u_*=0.5$}
    \psfrag{ustar0.6}[r][r]{\tiny $u_*=0.6$}
    \psfrag{ustar0.7}[r][r]{\tiny $u_*=0.7$}
    \includegraphics[width=0.95\columnwidth]{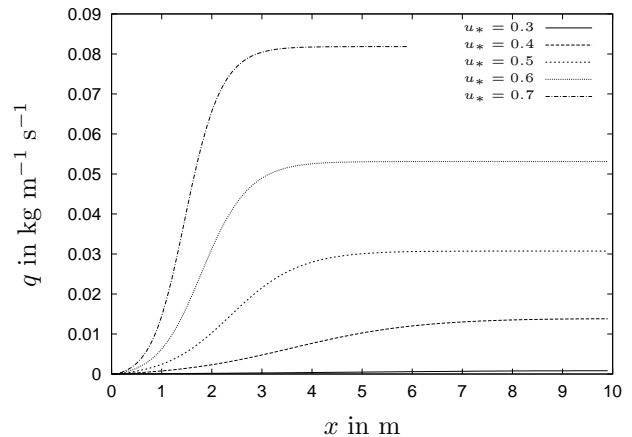}
    \vspace*{4mm}
    \caption{Numerical solution of the sand flux equation
      (\ref{eq:q_stationary}) for different shear velocities $u_*$.
      The model parameter $\gamma=0.2$ that defines the length and
      time of the saturation transients was chosen here so that
      saturation is reached between 1$\,$s and 2$\,$s. Due to the
      simplifications made with respect to the full model, the initial
      over--shoot is lost.}
    \label{fig:q_vs_x}
  \end{center}
\end{figure}

The saturated flux $q_s$ (\ref{eq:q_s_steady}) and the model
parameters $\alpha=0.35$ and $z_1=0.005\,$m have been discussed in
section \ref{sec:steady_state} and can be used unchanged in \Eq
(\ref{eq:q_stationary}). But with respect to the full model of section
\ref{sec:dyn} the value of $\gamma$ gets renormalized due
to the simplifications made in order to obtain \Eq
(\ref{eq:q_stationary}). The most obvious difference of the solution
of \Eq (\ref{eq:q_stationary}), depicted in Figure
\ref{fig:q_vs_x}, compared to the result of the full model in section
\ref{sec:dyn} is the missing over--shoot. The value of $\gamma=0.2$
had to be adapted in order to obtain saturation transients between
1$\,$s and 2$\,$s for typical values of $u_*$. The fact that the
saturation length $l_s$ increases for shear velocities near the
threshold is unchanged and shown in Figure \ref{fig:q_vs_x} and
\ref{fig:ls}.

\section{The Sand Flux on the Windward Side of a Dune}
\label{sec:application}

\begin{figure}[tb]
  \begin{center}
    \psfrag{wind}[r][r]{wind}
    \psfrag{stoss-side}{windward side}
    \psfrag{slip-face}{slip face}
    \psfrag{brink}{brink}
    \psfrag{horn}{horns}
    \includegraphics[width=0.95\columnwidth]{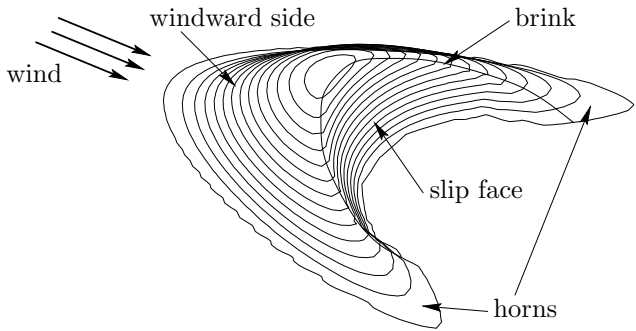}
    \caption{Sketch of a barchan dune.}
    \label{fig:sketch_of_barchan}
  \end{center}
\end{figure}

The mass conservation, \Eq (\ref{eq:dhdt}), and the saturated
sand flux relation by Lettau and Lettau, \Eq (\ref{eq:q_lettau}),
have been used many times to predict the evolution of a dune
\cite{Wippermann86,Zeman88,Stam97}.  The most successful work was done
by Wippermann and Gross \cite{Wippermann86}, but even they obtained an
unphysical deposition at the windward foot of the dune, which tended
to stretch the dune in length and to flatten it, finally. This was
avoided by an ad hoc smoothing operation, which led to a numerically
stable simulation by hiding the problem at the phase boundary
(bedrock/sand). The boundary problem is evident for an isolated dune
like a barchan sketched in Figure \ref{fig:sketch_of_barchan}. Recent
field measurements \cite{Wiggs96} of the wind speed and the sand flux
on the central slice of a barchan dune show a large discrepancy
between the measured sand flux at the windward foot and the sand flux
predicted by \Eq (\ref{eq:q_lettau}) for the measured wind speed.
The measured sand flux was a monotonously increasing function, whereas
the calculated sand fluxes decreased near the dune's foot due to the
depression of the wind velocity. A decrease of the flux is correlated
with deposition at the dune's foot leading to a flattening of the
dune.  Already from the measurements it is evident, that a saturated
flux law cannot be applied near a phase boundary, where the bed
changes suddenly from bedrock to sand. A further problem apears if we
integrate \Eq (\ref{eq:dhdt}) in time using a flux relation of
the form $q(u_*)$ that determines the sand influx
$q_{in}=q[u_*(x_{in})]$ and outflux $q_{out}=q[u_*(x_{out})]$ at the
boundaries. This might be true for the outflux, but not for the
influx, which depends normally on the upwind conditions and not on the
wind speed at the boundary. In particular, the influx of an isolated
dune in a dune field should depend on the outflux of several dunes
upwind. Even situations without influx are possible, e.g. if there is
vegetation around the dune.

\begin{figure}[t]
  \begin{center} 
    \psfrag{tau}[b][c]{$\tau$ in Pa}
    \psfrag{flux}[b][c]{$q$ in kg~m$^{-1}$~s$^{-1}$}
    \psfrag{height}[b][c]{$h$ in m}
    \psfrag{position}[t][c]{$x$ in m}
    \psfrag{Lettau}[l][l]{\tiny $q_L$ Lettau}
    \psfrag{SauermannXXXXXXXXXX}[l][l]{\tiny $q$, non--saturated}
    \psfrag{Sauermannsteady}[l][l]{\tiny $q_s$, saturated}
    \includegraphics[width=0.95\columnwidth]{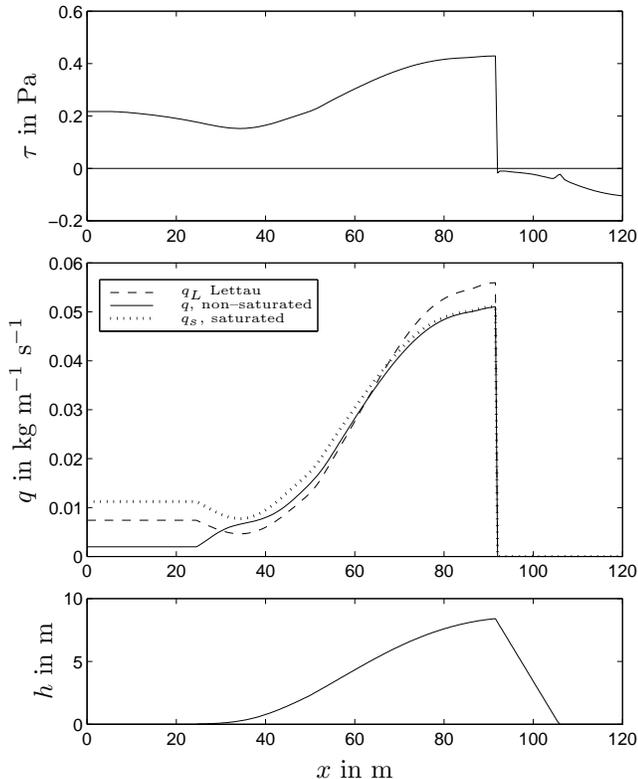}
    \vspace*{4mm}
    \caption{Top: shear stress $\tau$ of the air at the surface,
      calculated by a $k\epsilon$-turbulence model using FLUENT 5.
      Center: The sand flux according to the ``minimal model'',
      \Eq (\ref{eq:q_stationary}), the saturated sand flux $q_s$,
      \Eq (\ref{eq:q_s_steady}), and the sand flux $q_L$
      predicted by the Lettau and Lettau relation (\ref{eq:q_lettau})
      have been calculated using the shear stress depicted at the top.
      Bottom: Height profile of the symmetry--plane of a barchan.}
    \label{fig:flux_shear}
  \end{center}
\end{figure}

To elucidate this problem further, we calculated the shear stress
using FLUENT~5 \cite{Fluent5} with a $k\epsilon$-turbulence model for
the central profile, parallel to the wind direction, of a barchan
measured in Morocco. (For more details see dune no. 7 in
\cite{SauermannRognon2000}.) The exact calculation of the flow field
is not of importance for the following discussion, which only relies
on characteristic qualitative features such as the depression near the
dune's foot. For the calculation of the sand flux we assume bedrock up
to the dune's foot ($x_{in}$=25m), where erosion is not possible
($q$=constant). The sand flux over the bedrock is therefore equal to
the influx at the boundary $x_{in}$. The sand flux $q_L$ according to
the relation by Lettau and Lettau, \Eq (\ref{eq:q_lettau}), and
the solution of the saturated flux $q_s$, \Eq
(\ref{eq:q_s_steady}), predicted by our model are depicted in
Figure~\ref{fig:flux_shear}.  Both models exhibit the unphysical
deposition at the dune's foot described above. This problem is
resolved, when the ``minimal model'', \Eqs
(\ref{eq:q_stationary}), (\ref{eq:l_s_of_u_s}), and
(\ref{eq:q_s_steady}), is applied. In contrast to a saturated flux
law, the boundary conditions can freely be chosen in this model. Here,
we used a constant influx $q_{in}$, much smaller than the saturated
one, which represents the inter--dune sand flux. At the outflow
boundary we applied $\partial q / \partial x = 0$. The solution is
plotted in Figure~\ref{fig:flux_shear} in comparison with the
predictions of the simple flux relations. Due to the saturation
transients we obtain a monotonously increasing flux and therefore no
deposition of sand at the dune's foot. Finally, we want to point out
that the flux on the entire windward side is never fully saturated due
to the continuously increasing shear stress. However, away from the
boundary it is always close to the saturated value (compare $q$ and
$q_s$ in Fig. \ref{fig:flux_shear}), which justifies the
simplifications made in the ``minimal model''.

\section{Conclusion and Outlook}

We derived a phenomenological sand transport model that reproduces the
equilibrium sand fluxes measured in wind tunnels.  The predicted
evolution of the sand flux with time has the same qualitative average
behavior as the sand flux calculated with saltation models working on
the microscopic grain scale. Finally, we proposed a ``minimal model''
that can be used as an efficient tool for geomorphological
applications, such as the formationn and migration of dunes.
Furthermore, we showed that this model extends in a general way common
saturated sand flux relation to a model that incorporates saturation
transients.

The phenomenological parameters for the saltation model, which are an
effective restitution coefficient, a reference height within the
saltation layer, and a saturation length have been estimated by
comparison with measurements.

Using these parameters we applied our model to a geomorphological
problem and calculated the sand flux over a dune.  We emphasized the
importance of a non--equilibrium flux relation for the correct
modeling of phase boundaries, e.g. bedrock/sand, as they naturally
occur for isolated dunes. Furthermore, we have shown that the model
predicts saturation transients to persist over the entire windward
side of a dune, where the shear stress increases from the foot to the
brink. Therefore, we claim that the saturation length defines a length
scale that is important for dune morphology in general. The
investigation of the implications of saturation transients for dune
formation will given in Ref. \cite{unpub:KroySauermann2001}. In
particular, the question of shape differences between small and large
dunes or the minimum size for slip--face formation are of interest and
are discussed there. The extension of the model to three dimensions
and inclined surfaces, which is necessary for applications to
arbitrary terrains will be the subject of futur work.

\section{Acknowledgement}

We acknowledge the support of this work by the Deutsche
Forschungsgemeinschaft (DFG) under contract No. HE 2731/1-1.
Furthermore, we thank J. Soares Andrade Jr. for many fruitful
discussions.

\bibliography{journals,dune,books,unpublished}

\end{multicols}
\end{document}